\documentstyle[aps,prl]{revtex}
\begin{document}
%\makeatletter
%\@addtoreset{equation}{section}
\renewcommand{\theequation}{\thesection.\arabic{equation}}
\makeatother

\newcommand{\be}{\begin{equation}}
\newcommand{\ee}{\end{equation}}
\newcommand{\<}{\langle}
\renewcommand{\>}{\rangle}
\newcommand{\reff}[1]{(\ref{#1})}

\title{Finite temperature QCD in the quark-composites approach}  
\author{ F. Palumbo{ ~\thanks
       {This work has been partially 
  supported by EEC under TMR contract ERB FMRX-CT96-0045}   \\}}
 \address{ 
     {\it INFN -- Laboratori Nazionali di Frascati}  \\
  {\it P.~O.~Box 13, I-00044 Frascati, ITALIA}          \\
  { Internet: {\tt palumbof@lnf.infn.it}}     
   }
\maketitle

\thispagestyle{empty}   % Suppress page number on front page.

\begin{abstract}
 We investigate QCD at finite temperature in the quark composites approach, which is based on the use
of quark composites with hadronic quantum numbers as fundamental variables. We find that chiral symmetry
restoration and quark deconfinement are one and the same first order phase transition, whose critical 
temperature, in a one loop approximation, is $T= 2\sqrt{\Omega}  \rho^{-2} m_{\pi}$, where $m_{\pi}$ is the pion
mass, $\Omega=24$ the number of up and down quark components, and $\rho$ a parameter of order 1 whose precise
value can be determined by the study of the pion-pion interaction.  \end{abstract}

\clearpage

\twocolumn

The idea that a spontaneously broken symmetry can be restored at sufficiently high temperature~\cite{Lind}
has found a natural application in QCD. Since quantitative investigations of this problem rely so far on numerical
simulations, it is not easy to establish whether the chiral symmetry restoration is a first or second order
phase transition, a feature which according to many authors depends on the number of flavors~\cite{Kars}. The
situation is somewhat complicated by the possible interplay between the restoration of chiral symmetry and quark
deconfinement, which in a recent paper have been found to coincide~\cite{Kars1}.

We study QCD at finite temperature in the quark composites approach~\cite{Palu} with two flavors. We find that
chiral  symmetry restoration and quark deconfinement are indeed one and the same first order phase transition.
Although the validity of our resul is restricted to two flavors, the reason why the transition is first order
does not seem to depend on the number of flavors. We find the critical temperature 
\be 
T= { 2\sqrt{\Omega} \over \rho^2}\; m_{\pi}. \label{tem} 
\ee
In the above equation $m_{\pi}$ is the pion mass, $\Omega=24$ is the number of up and down quark components, and
$\rho$ is a parameter of order 1 whose precise value can be determined by the study of the pion-pion
interaction. 

While the conclusion about the nature of the phase transition is unavoidable in the present context, we cannot
say anything about the accuracy of the value  of the critical temperature  without an estimate of its
corrections. Here it could help a numerical simulation done using the action of the quark composites
approach. Indeed this approach is based on the idea that if a significant part of the binding mechanism of
hadrons is accounted for by the use of quark composites with hadronic quantum numbers as fundamental variables,
the residual interaction can be sufficiently weak for a perturbative treatment. If this is true the present
formulation should also make it easier the numerical evaluation of the quark determinant.

 The idea at the basis of the quark composites approach is realized by assuming the composites with the quantum
numbers of the nucleons as new integration variables in the Berezin integral which defines the partition
function~\cite{DeFr}, and replacing the composites with the quantum numbers of the chiral mesons by auxiliary
fields~\cite{Cara}. Then irrelevant operators which provide the kinetic terms for the composites are added to the
standard action, and the quark action is treated as a perturbation. The expansion parameter cannot  obviously be
the gauge coupling constant, but it is instead a dimensionless constant appearing in the definition of the
composites. In conclusion the perturbative expansion is formulated in terms of phenomenological fields, whose
effective coupling constants, however, are given to any order by integrals, depending on the order, over the
gluon and quark fields. The gluon fields can be treated, according to the dynamical regime, perturbatively or non
perturbatively wr to the gauge coupling. It is an intrinsic, fundamental feature of the approach that the quarks,
as a consequence of the spontaneous chiral symmetry breaking are nonpropagating particles which appear only in
virtual states: In other words their confinement is built in already at the perturbative level, unlike the
standard perturbation theory where the quarks appear as physical particles whose confinement is intrinsically
nonperturbative.

The framework of our approach appears most convenient for the study of the phase transitions associated to chiral
symmetry restoration and quark deconfinement. In this paper we investigate QCD at finite temperature and zero
barion density, a purpose for which it is sufficient to introduce only the chiral composites. We adopt a
regularization on a euclidean lattice because it seems natural dealing with composites, and it allows us to
handle also the nonperturbative regime of the gluons. We assume the modified partition function 

\begin{equation}
Z = \int [dU] \int [d\overline{\lambda} d\lambda] \exp[-S_{YM} -S_{C} -S_Q], 
\end{equation}
where $S_{YM}$ is the Yang-Mills  action, $S_Q$ is the action of the quark fields and $S_C$ is an
irrelevant operator which provides the kinetic terms for the quark composites with the quantum numbers of the
chiral mesons. $\lambda^a_{\tau \alpha}$ is the quark field with color, isospin and Dirac indices $a$, $\tau$ and
$\alpha$ respectively, and the gluon field is  associated to the link variables $U_{\mu}$. 

All the elementary fields live in an euclidean lattice of spacing $a$, whose sites are identified by
fourvectors $x$ of spatial components $x_k=0,...N$ and time component $t=0,...N_t$, and satisfy periodic
boundary conditions, with the exception of the quark fields which are antiperiodic in time: $\lambda(x)=
\lambda(x+Ne_k)= - \lambda(x+ N_t e_t)$,  $e_k$ and $e_t$ being the unit vectors in the $k$ and time directions.
The temperature is $T= (N_t a)^{-1}$.
 
The quark action, with the notation 
\be
 (f,g)  =  a^4 \sum_x f(x) g(x),
\ee
can be written
\be
S_Q= (\overline{\lambda}, Q \lambda ).
\ee
The quark wave operator is
\be
Q_{x,y}  =  m_Q 
 \delta_{xy} +{1\over 2a}\sum_{\mu}  
  \gamma_{\mu} U_{\mu}(x) \delta_{x+e_{\mu},y}.
\ee
We adopt the standard conventions
\begin{eqnarray}
 & & \mu  \in  \{ -4,\ldots,4\},\;\;\;e_{-\mu}=-e_{\mu}, 
\nonumber\\
 & & \gamma_{-\mu}  =  - \gamma_{\mu},
\nonumber\\
 & & U_{-\mu}(x)  =  U^+_{\mu}(x-e_{\mu}).
\end{eqnarray}
and omit the Wilson term for a reason to be explained later.

The chiral composites are the pions and the sigma
\be 
  \pi_h = i\,k_{\pi}\,a^{2}\,\overline{\lambda} \gamma_5  \tau_h  \lambda, \;\;\;  
\sigma = k_{\pi}\,a^{2}\,\overline{\lambda} \lambda. 
\ee
 $\gamma_5$ is assumed hermitean and the $\tau_{k}$'s are the Pauli matrices. An arbitrary factor of the 
dimension (length$)^2$, necessary to give the chiral composites the dimension of a scalar field, has been
written for convenience in the form $ a^2 k_{\pi}$. The composite action must be $O(4)$ invariant, so that with
the inclusion of a linear breaking term it must have the form 
\begin{equation}
S_{C} =  \left[ {1 \over 2} (\vec{\pi},C\vec{\pi})_c  + {1 \over 2}
(\sigma,C\sigma)  - { 1 \over a^2} \left(  \sqrt {\Omega}m,
\sigma \right) \right], \label{actionC2}   
\end{equation}
with $\vec{\pi} \cdot \vec{\pi} = \pi_1^2+ \pi_2^2  +\pi_3^2$. The factor $\sqrt \Omega$ (we remind that 
$\Omega$ is the number of quark components), has been introduced for later convenience. We choose~\cite{Cara} 
the wave operator   
\be
 C = a^{-2}{\rho^4 \over a^2 {\Box - \rho^2}}, 
\ee
where $\Box$ is the laplacian on the lattice. The irrelevance of $S_C$ requires that the parameter $\rho$
 be independent on the lattice spacing and that $k_{\pi}$ do not diverge in the continuum limit.
These constraints, with the dependence on $a$ assumed below for the breaking parameter $m$, ensures also the
irrelevance of the chiral symmetry breaking term.

We will now sketch the derivation of a perturbative expansion for the partition function of QCD. We start by using
the Stratonovich-Hubbard transformation~\cite{Hub} to replace the chiral composites by auxiliary fields 
\be 
 \exp[-S_C]  =  \left[{d\vec{\chi}\over\sqrt{2 \pi}}\right]  
   \left[{d\phi\over\sqrt{2 \pi}}\right] \exp [-S_{\chi}]
\bigtriangleup \exp    \left(\overline{\lambda},{1 \over a} D \lambda \right) ,  
\ee
where
\begin{eqnarray}
 S_{\chi} & = & -{1\over 2 } \rho^4 \left[ (\vec{\chi},(a^4 C)^{-1}\vec{\chi})  
+ (\phi,(a^4 C)^{-1} \phi) \right] 
\nonumber\\
 & & - { \Omega \over 2} \sum_x \ln \left\{ a^2 k_{\pi}^2 \left[ \left(
    \sqrt{ \Omega} m + \rho^2 \phi(x) 
   \right)^{2} +  \rho^4 \vec{\chi}(x)^{2} \right] \right\}, 
 \nonumber\\
\bigtriangleup &=& \prod_x (\det D(x))^{-1} ,
\nonumber\\
D &= & a  k_{\pi}\left[\rho^2\phi + \sqrt \Omega m + i \rho^2  \gamma_5 \vec{\tau} \cdot \vec{\chi}\right]. 
\end{eqnarray}
The partition function can then be written
\begin{eqnarray} 
   Z & = &\int [dU] \exp[-S_{YM}] \left[{d\vec{\chi}\over\sqrt{2 \pi}}\right] 
  \left[{d\phi \over\sqrt{2 \pi}}\right] \exp[-S_{\chi}]
\nonumber\\
   & &\bigtriangleup \int [d \bar{\lambda} d\lambda]
 \exp \left( \overline{\lambda},({1 \over a} D + Q )\lambda \right) .   \label{Zsim}
\end{eqnarray}

The minimum of $S_{\chi}$ occurs at $\vec{\chi}=0$, but $\phi \neq 0$, implying a breaking of the chiral symmetry.
Since $\Omega$ is a rather large number we can  apply the saddle-point method and perform an expansion in inverse
powers of this parameter~\cite{Cara}. The validity of this expansion is subject~\cite{Palu1} to the condition
$\rho \sim 1$ mentioned in the introduction. $S_{\chi}$ contains a quadratic part plus interactions which vanish
in the  continuum limit (but have finite quantum effects). The masses associated to the fields $\phi, \chi$ and 
the mass of the quarks as functions of $\phi$ are 
\begin{eqnarray}
  m^2_+ & = &{\rho^2 \over a^2}\left\{ 1+ \left({1\over \sqrt{\Omega}}\rho a \phi+ 
{ 1\over \rho} a m \right)^{-2} \right\}
\nonumber\\
 m^2_- &=& {\rho^2 \over a^2}\left\{ 1- \left({1\over \sqrt{\Omega}}\rho a \phi+ { 1\over \rho} a m \right)^{-2}
\right\},
\nonumber\\
M_Q &= & k_{\pi}(\rho^2 \phi+ \sqrt{\Omega}m) + m_Q.
\end{eqnarray}
The minimum of $S_{\chi}$ wr to $\phi$ occurs at 
 \be
\phi=\overline{\phi} \sim { \sqrt {\Omega} \over a\; \rho}( 1- \xi ),
\ee
where
\be 
\xi= { 1\over 2 \rho} a m.
\ee
The mass of the pion coincides with $m_-$. To have $m_- = m_{\pi}$ we must assume the breaking parameter
\be
  m= { 1\over \rho} a\; m_{\pi}^2 .
\ee
The mass of the $\sigma$, which is equal to $m_+$, diverges in the continuum limit.
 
The quarks  have a mass  $M_Q \sim \sqrt{\Omega} \rho k_{\pi}/ a $, which seems also to diverge in the continuum
limit. Whether this is actually true depends on the behaviuor of $k_{\pi}$ with the lattice spacing. An
exemple of the possible need of a scaling of $k_{\pi}$ with $a$ is provided by the expression of
the pion-nucleon coupling constant~\cite{Palu1}. But irrespective of this dependence the quarks are
nonpropagating at the perturbative level. Since their mass is of order $\sqrt{\Omega}$, the saddle point
expansion results to be a hopping expansion as far as the quarks are concerned, and as a consequence, the quark
correlator connecting two sites which are $n$ lattice spacings apart from one another gets the first
nonvanishing contribution  at an order of the expansion not smaller than $n$: The quarks are never produced to
any finite order. In this sense we can say that they are confined at the perturbative level. 

In connection with the nature of the phase transition, it is essential to observe that the above result is solely
due to the spontaneous chiral symmetry breaking, in whose absence there would be no saddle point expansion, and
does not depend on the explicit breaking term. Since the quark confinement is due to the spontaneous chiral
symmetry breaking, the chiral symmetry restoration coincides with the quark deconfinement: They are one and the
same phase transition.

We also note that since in the confining phase the quarks have no poles whatsoever, it seems that we do not have
to worry about the spurious ones. In any case we can safely assume the quark Wilson parameter $r_q$  of order $1/
\sqrt{\Omega}$, so that we can neglect the quark Wilson term to leading order in our expansion.

Introducing the fluctuation of the $\phi$-field
\be
\theta=  \phi-\overline{\phi},
\ee
we arrive at the desired expansion of the partition function of QCD in terms of the auxiliary fields. Since only 
terms with an even power of $S_Q$  can contribute, we have   
\begin{eqnarray}
 Z & = &\int [dU] \exp[-S_{YM}] \left[{d\vec{\chi}\over\sqrt{2 \pi}}\right] 
  \left[{d\phi \over\sqrt{2 \pi}}\right] \exp[-S_{\chi}]\bigtriangleup 
  \nonumber\\
 & &  \int [d\bar{\lambda} d\lambda]\sum_{r=0}^{\infty}{1 \over (2r)!}  
  ( S_Q)^{2r} \exp  \left(\overline\lambda,{1 \over a} D \lambda \right)   = 
\sum_{r=0}^{\infty} Z_r.
 \end{eqnarray}
This is an expansion in inverse powers of $k_{\pi}$, because obviously $ Z_r \sim k_{\pi}^{-2r}$. Each term
$Z_r$ is a function of $\Omega$ which can be given a series in $\Omega^{-1}$. The study of the pion-pion
interaction determines the value of $\rho$ and the scaling of $k_{\pi}$ with the lattice spacing~\cite{Cara1}.
Note that we do not need to treat the gluon field perturbatively: If and where  this can possibly be done
remains here an open question.

Finally we  evaluate the critical temperature in a one loop approximation, by the condition that the effective
potential  be a minimum for vanishing pion mass, namely for  
\be 
\phi=\overline{\phi} - { \sqrt{\Omega} \over 2 \rho^3}  a \;m_{\pi}^2. 
\ee
For this value of $\phi$ the mass of the $\sigma$ remains divergent in the continuum limit: the chiral
symmetry restoration-quark deconfinement is a first order phase transition.

It is  convenient to introduce the field $\varphi$
\be
\overline{\phi}(T) = \overline{\phi} + a \varphi,
\ee 
because the pion mass vanishes for a finite value of it
\be
\overline{\varphi}= - { \sqrt{\Omega} \over 2 \rho^3} m_{\pi}^2.
\ee
Then, since $\varphi$ has dimension 2, we can confine ourselves to approximations quadratic in this field.
 Let us separate the effective potential in its various parts
\be
V = V_0 + V_1 + V_T.
\ee

$V_0$ is the classical potential
\begin{eqnarray}
V_0 &=& {1\over 2 a^2}\rho^2 \phi^2  - \Omega { 1\over a^4} \ln \left( \sqrt{\Omega} am + \rho^2 a \phi
\right)
\nonumber\\
 & & \sim \mbox{const} + \rho^2 \varphi^2.
\end{eqnarray}
$V_1$ is the one loop contribution at zero temperature.  Its quadratic approximation, under the usual
normalization conditions  
\be 
{\partial \over \partial \varphi} V_1= {\partial^2 \over \partial \varphi^2} V_1=0,
\ee
 must vanish identically. This has been checked by an explicit evaluation of the counterms $\delta m$
and $\delta \rho$ necessary to cancel out the one loop divergences. The calculation has been performed for 
simplicity by considering the action $S_{\chi}$ in a continuous euclidean space, whith a cutoff $1 /a$. Then  
the contribution coming from the counterterms is
\begin{eqnarray}
\overline{V}_1 &=& {\sqrt{\Omega} \over b^2 a^2}\left( a \delta m + { \delta \rho^2 \over \rho b } \right) \varphi
\nonumber\\
 & & + \left[ \left( {1 \over 2} + { 1 \over b^2} \right) \delta \rho^2 - { \rho \over b^3} \left( a \delta m + 
{ \delta \rho^2 \over \rho b } \right) \right] \varphi^2.
\end{eqnarray}
The total zero temperature one loop potential is
\be 
V_1 = \overline{V}_1 + F(m_+^2) + 3 F(m_-^2),
\ee
where
\be
F(m^2) = { 1\over 16 ( 2\pi)^2} \left[ { 1\over 2} m^4 + 4 a^{-2} m^2 - m^4 \ln { 4 \over a^2 m^2} \right].
\ee
The approximations to $m^2_{\pm}$, quadratic in the field $\varphi$, are
\begin{eqnarray}
m^2_{+} &=& {2 \rho^2 \over a^2}\left( 1 - \xi - { \rho  \over \sqrt{\Omega}} a^2 \varphi
 - {3\rho^2  \over 2\Omega} a^4 \varphi^2  \right)  
\nonumber\\
m^2_{-} &=&{2 \rho^2 \over a^2}\left( \xi + { \rho  \over \sqrt{\Omega}}  a^2 \varphi + 
{3\rho^2  \over 2 \Omega} a^4\varphi^2  \right).
\end{eqnarray}
>From the cancellation of the divergent terms we get
\begin{eqnarray}
\delta \rho^2 &=& { 1\over (2\pi)^2}{ \rho^6\over \Omega} \left[ -{3\over 2} +{1 \over 3\rho^2} -{1 \over 3}
\ln{\rho^2\over 2} -{ 1\over 2} \ln { \rho^2 \xi \over 2}\right]
 \nonumber\\
\delta m &=& { 1\over a}{ 1\over (2\pi)^2}{ \rho^5\over \Omega}\left[ 2 -{ 4 \over 3\rho^2} 
 +{5 \over 6} \ln{\rho^2\over 2}  +{ 1\over 2} \ln { \rho^2 \xi \over 2} \right],
\end{eqnarray}
which make $V_1$ identically zero.

 Finally  $V_T$ is the temperature dependent one loop part
\be
V_T = V^{+}(T) + 3V^{-}(T) - V^{+}(0) + 3V^{-}(0),
\ee
where 
\begin{eqnarray}
V^{\pm}(T) &=& { T \over 2 (2 \pi)^3} \; \int_{-{\pi \over a}}^{+{\pi \over a}}d^3 p 
\nonumber\\
& &\sum_n  \ln\left(  a^{-2} \sin^2 \left( { 2\pi \over N_t}n \right) + p^2 + m_{\pm}^2  \right).
\end{eqnarray}
$V^{\pm}$ has been evaluated in~\cite{Lind} for $T \gg m$. Since we do not need its full expression, 
but only  its derivative, we can avoid such limitation.
Now
\begin{eqnarray}
{\partial \over \partial m^2_{\pm}} V^{\pm} & =& { T \over 2(2 \pi)^3} \; \int_{-{\pi \over a}}^{+{\pi \over a}}
d^3 p 
\nonumber\\
  & & \sum_n  \left( a^{-2} \sin^2 \left( { 2\pi \over N_t}n \right) + p^2 + m_{\pm}^2  \right)^{-1}.
\end{eqnarray}
Taking the limit $a \rightarrow 0$ at finite $T$ we get
\begin{eqnarray}
{\partial \over \partial m^2_{\pm}} V^{\pm} & =& { T \over 2(2 \pi)^3} \; \int_{-\infty}^{+\infty}
d^3 p 
\nonumber\\
  & & \sum_n  \left(  \left(  2\pi n T\right)^2 + p^2 + m_{\pm}^2  \right)^{-1}.
\end{eqnarray}
 We see that the $\sigma$ does not contribute because of its
divergent mass, while the pion contribution for vanishing pion mass is
\begin{eqnarray}
{\partial \over \partial m^2_{-}} & & \left[ V^{-}(T) -  V^{-}(0) \right]  ={ 1 \over 2 (2 \pi)^3} \;
\int_{-\infty}^{+\infty} d^3 p 
\nonumber\\
  & & { 1\over p}\left[ \exp (p^2 /T) -1 \right]^{-1}={ 1 \over 24} T^2.
\end{eqnarray}
In conclusion
\be
{\partial \over \partial \varphi} V = 2\rho^2 \varphi + {\rho^3 \over 4 \sqrt{\Omega}}T^2.
\ee
>From the condition that the effective potential be a minimum for vanishing pion mass, namely for
$\varphi = \overline{\varphi}$ we get the critical temperature 
\be 
T= { 2\sqrt{\Omega} \over \rho^2}\; m_{\pi}. 
\ee
To have an estimate of the accuracy of this result one should evaluate the two loop contribution to
the effective potential taking into account the quark contribution in the expansion in $k_{\pi}^{-1}$. 
Another possibility is to use the formulation \reff{Zsim} in a numerical simulation. This might have
the advantage of an easier evaluation the quark determinant and of the guidance of the above
result.

We conclude by observing that our approach appears promising also in the study of the phase transition at high
baryon density~\cite{QCD} because, unlike the standard way, where the chemical potential acts on the quarks to
allow the integration over the latters to be performed analytically, we can couple it directly to the nucleons,
with a significant simplification.

%\end{twocolumn}
\end{document}